\documentclass[manuscript]{acmart}

\copyrightyear{2026}
\acmYear{2026}
\setcopyright{cc}
\setcctype{by}
\acmConference[CUI '26]{ACM Conversational User Interfaces 2026}{July 21--24, 2026}{Bremen, Germany}
\acmBooktitle{ACM Conversational User Interfaces 2026 (CUI '26), July 21--24, 2026, Bremen, Germany}
\acmDOI{10.1145/3816046.3816203}
\acmISBN{979-8-4007-2741-2/2026/07}

\usepackage{color,soul}
\usepackage{multirow}

\newif\ifstatus
\statustrue
\begin{document}

\title{Beliefs and Misconceptions around Integrated Conversational AI}

\author{William Seymour}
\email{william.seymour@kcl.ac.uk}
\orcid{0000-0002-0256-6740}
\affiliation{%
  \institution{King's College London}
  \streetaddress{Bush House, 30 Aldwych}
  \city{London}
  \country{UK}
  \postcode{WC2B 4BG}
}

\author{Adam Jenkins}
\email{adam.jenkins@kcl.ac.uk}
\orcid{0000-0001-7865-0087}
\affiliation{%
  \institution{King's College London}
  \streetaddress{Bush House, 30 Aldwych}
  \city{London}
  \country{UK}
  \postcode{WC2B 4BG}
}

\author{Mark Cot\'{e}}
\email{mark.cote@kcl.ac.uk}
\orcid{0000-0001-6359-1627}
\affiliation{%
  \institution{King's College London}
  \streetaddress{Chesham Building, Strand}
  \city{London}
  \country{UK}
  \postcode{WC2R 2LS}
}

\author{Jose Such}
\email{jose.such@csic.es}
\affiliation{%
  \institution{Ingenio (CSIC-Universtiat Politècnica de València)}
  \streetaddress{Cami de Vera s/n}
  \city{València}
  \country{Spain}
  \postcode{46022}
}

\renewcommand{\shortauthors}{Seymour and Jenkins, et al.}

\begin{abstract}
LLM-driven conversational AI is beginning to disappear into the background, shifting from something used directly towards something increasingly integrated into existing workflows. In the process, markers of origin and training are smoothed away as LLMs become commodified in the eyes of users. We explore how people approach using a web browser with conversational AI built in, focusing on how they develop their understanding and determine whether to trust its outputs. We conducted a study where 20 participants used the Copilot AI features in Microsoft Edge to conduct information retrieval and planning tasks. Participants relied on a combination of existing perceptions of LLMs and internet search, tracing the effect of beliefs about how Copilot generated answers on prompting strategies. The inclusion of citations increased the trustworthiness of answers without participants feeling the need to be check them, with participants often reaching for the same information sources as the CAI when fact-checking.
\end{abstract}




\maketitle

\section{Introduction}
Having initially entered the mainstream through devices such as Alexa and Siri, conversational AI (CAI) has exploded in popularity through the development of Large Language Models (LLMs). Successive iterations of increasingly sophisticated LLMs have offered much more human-like conversation than that which was previously possible---the Alexa Prize SocialBot `Grand Challenge' criterion of coherently conversing on popular topics and news events for 20 minutes almost seems trivial only a few years on. Initially, the technology was available as a standalone product, with ChatGPT being the most recognized. However, as the novelty begins to wear off there is increasing demand for this technology to be integrated into existing workflows~\cite{weber2024large}; it now feels normal to collaborate with a CAI agent to achieve a range of everyday tasks, from AI summaries in Google searches to GPT4 powered Duolingo chats.

But this transition is not without its drawbacks, primarily that companies seem far more interested than users or employees in adopting LLMs~\cite{joannapartridge2026, adverse}. There are also a lack of opportunities to acquire the skills and norms required to use LLMs effectively and appropriately~\cite{johnnyCant}, partially resulting from early responses where (particularly academic) organisations created policies seeking to curtail AI use. In practice, the tools themselves defined their own norms by making LLM use the easiest or default option for many tasks; what does a policy forbidding the use of AI in an assignment mean when both Word and Overleaf have built-in AI text generation and editing features? The vast amounts of knowledge and computational power required to train foundation models has meant that a small number of these models are trained, tuned, and then rebadged for deployment into a seemingly endless array of apps and services. In the process of becoming ubiquitous, markers of origin, training, and version are obscured as LLMs become commodified in the eyes of the public~\cite{liesBeneath}; Duolingo users are likely unaware or apathetic about which LLM model the app uses, but this matters given that we have seen how models can be put to varying business and political ends (e.g. Grok).

To help address these issues, this paper explores how users approach using new conversational AI models, focusing on how they develop understandings of strengths, weaknesses, and risks. We present the results of a user study where 20 participants were asked to complete an information retrieval and a planning task using the Copilot CAI integrated in the Microsoft Edge web browser. We captured their initial responses and perceptions to the new tool before asking them to reflect on their strategies and the quality of the responses they received. Through our analysis we answer the following research questions:

\begin{enumerate}
    \item[RQ1] What beliefs guide user experiences with new LLMs?
    \item[RQ2] How do users navigate issues of trust and reliability with a new LLM?
\end{enumerate}

\section{Background}
%As the range of applications grows, researchers have also explored on the use of CAI in healthcare~\cite{zhan2024healthcare, 10.1145/3584931.3606993, 10.1145/3678884.3681858, 10.1145/3678884.3681921} as well as a range of other specific use cases for the technology~\cite{10.1145/3678884.3681909, 10.1145/3584931.3606965} and interface design~\cite{10.1145/3686902}.

\subsection{Perceptions of Web Search}
Prior work on web search has identified a range of concepts that were believed to influence search engine result pages, including popularity, wording, commercial interests, and personalisation, and found that the limited explanations given to users by search engines did not appear to influence the construction of mental models~\cite{10.1145/3371390}. Identified targets for search engine explanations were payment (in relation to search rankings), results filtering, alongside the recency, diversity, and credibility of results. Relatedly, the need for transparency in search and recommender systems is often unquestioned, with researchers believing that providing system explanations may improve users' perceptions and resolve trust issues. However, work exploring \textit{if} explanations are necessary and \textit{when} such details should be provided suggests that people value explanations when performing complex or critical tasks and require no explanations when the tasks are simple or provide an unambiguous answer~\cite{10.1145/3613904.3642059}. \citeauthor{10.1145/3613904.3642059} report that participants expressed a preference for the ability to challenge search results and the need for actionable explanations~\cite{10.1145/3613904.3642059}.

\subsection{Perceptions of AI and CAI}
There has been a sustained interest in exploring how people perceive CAI agents and their many applications, initially with voice assistants and more recently with LLMs. Prior work has shown the oversized influence of non-functional aspects of CAI design. Initial expectations tend to anchor subsequent experiences, with unmet high expectations leading to negative perceptions and reduced trust (and vice versa)~\cite{GRIMES2021113515}. The embodiment of voice-based CAIs pushes people towards perceptions of specialism rather than being all-purpose~\cite{10.1145/3411840}. This is likely to be related to issues of discoverability in voice interfaces, leading to simpler mental models that do not develop over time~\cite{10.1145/3570945.3607335, 10.1145/3264901}. This occurs because unlike graphical interfaces which can efficiently display suggested commands or follow-up actions, long conversation turns can be difficult to parse: ``any task where getting information that is dense or requires verbose explanation [..] is one where a customer probably won’t prefer to speak to an Alexa skill to accomplish it''.\footnote{\url{https://developer.amazon.com/en-US/alexa/alexa-haus/design-principles}} Other research has shown the effect of how smart and AI-driven systems are packaged and explained to users ~\cite{10.1145/3132031, 10.1145/3411840}, such as whether people's conceptualisations focus on the current interaction only, the current and previous interactions, or on the information the AI agent can access and how it uses it~\cite{10.1145/3313831.3376316}. Follow-up work has applied Item Response Theory to compare users' understandings of AI agents, themselves, and human collaborators~\cite{10.1145/3593013.3594111}.

Experimental data shows that participants tended to overestimate the abilities of the AI agents when compared to themselves, with a `general intelligence bias' leading to a view of AI agents as a single, unified intelligence even when presented with evidence suggesting otherwise~\cite{10.1145/3593013.3594111}. The exception to this is when initial experiences demonstrate system weaknesses, which leads to subsequent underestimation of a model's performance~\cite{10.1145/3397481.3450639}. There has also been some work on applying dual process theory to understandings of AI systems. Dual process theory says that human decision-making can be divided into two processes. Type 1 describes fast, heuristic cognitive processes and type 2 slower, effortful thinking. \citeauthor{10.1145/3686912} found that forcing type 2 thinking about AI systems helped users to more accurately understand what knowledge the system used and how~\cite{10.1145/3686912}. More specifically, we know that end users tend to struggle in similar ways as with end user programming when prompting LLMs~\cite{johnnyCant} and that overgeneralisation is also an issue, with a tendency for people to make binary decisions about whether an LLM can or cannot do something.

Several studies have explored trust around LLM use. Analysis of Bing Chat shows a correlation between trust and intention to use, with the aspects of the CAI rated lowest being information quality, trust, and risk perception~\cite{10417910}. In CAIs more generally it appears that in addition to information quality, trust is also moderated by functionality (i.e. correct but unusable answers lead to distrust) and predictability~\cite{Lee09102024}. A related body of work has begun to unpick the complex relationship between perceptions and understandings of CAI and subsequent trust in them~\cite{10.2196/47184}. At the same time, it has been observed that people automatically expect human-like interfaces to be more capable---that is, better at tasks suited to human intelligence---than those that are more machine-like~\cite{10.1145/3098279.3098539, 10.1145/2858036.2858288}). An apt framing might be that of folk psychology; people are generally good at predicting and attributing mental states in others~\cite{stich2003folk}, so when presented with a novel interface that converses like a human might treat it as such when deciding how to interact with it (e.g. when prompting LLMs, non-experts have been found to apply expectations from human-human interactions~\cite{johnnyCant}). Folk psychology is inherently messy---we know that predictions about what others think or are going to do may be wrong---but they do provide a starting point from which we can quickly adapt to a new interlocutor (much like Grice's cooperative principle).\footnote{\url{https://en.wikipedia.org/wiki/Cooperative_principle}}

%Given the impact of mental models on user interactions and perceptions of AI, research has attempted to address this by looking for explanations to provide insight into how AI systems work. Social sharing and validating shared mental models has also proven highly effective, with cycles of validation and re-sharing improving model quality and discounting inaccurate models~\cite{10.1145/3686912}.

\subsection{Explainable AI (XAI)}
Work on voice assistants suggests that different types of explanation may have different effects; while privacy explanations appear to mitigate privacy concerns, trust-based explanations appear to increase concerns about trust~\cite{10.1145/3579497}. Similarly, \citeauthor{10.1145/3581641.3584088} generated system-focused and social-focused explanations for a natural language system, the former detailing how a system handles inputs and generates responses and the latter how other users created their inputs to complete tasks successfully~\cite{10.1145/3581641.3584088}. Results indicated that social explanations encouraged more editing of inputs by participants and resulted in the successful completion of tasks when compared to system-focused explanations; however, neither showed marked improvement in users' mental models of the system in general. Interestingly, while explanations increase trust in CAI systems and user self-confidence, this seems to happen even when those explanations are incorrect~\cite{10.1145/3686922}, and students tend to have high confidence when using a CAI for learning even when task failure rates are high~\cite{10.1145/3617367}.

%Rather than using a single hub for information, people tend to trust different sources for different types of information; while Google is often the default choice, specific content types often lead people to other sources (like ideas from ChatGPT, definitions from Wikipedia, or entertainment from YouTube)~\cite{10.1145/3613905.3650862}.

%\subsection{The Social Nature of Conversational Interfaces}
%Early work on conversational interfaces suggested that computers are social actors in similar ways to people, causing us to apply social rules when using conversational interfaces even though they are clearly not human~\cite{nass1994computers}. This includes gendering interfaces/drawing on gender stereotypes~\cite{nass1994computers} and engaging in reciprocal information sharing as a social routine~\cite{moon2000intimate}. There has been debate over how these behaviours represent heuristics developed in lieu of accurate mental models; is saying `thank you' to ChatGPT the mindless execution of a social script or evidence of importing an existing mental model for a novel technology? While measures of human relationships may still hold with voice assistants~\cite{10.1145/3479515}, phenomena like the use of gendered pronouns do not appear particularly ingrained, with people switching between options and mostly using impersonal pronouns~\cite{10.1145/3359316, 10.1145/3027063.3053246}.

\subsection{Over-reliance, Bias, and Misinformation}
There have been concerns about over-reliance on LLMs, particularly in education. This echoes ongoing research on automation bias, ``the tendency to use automated cues as a heuristic replacement for vigilant information seeking and processing''~\cite{mosier2018human}, where there is some evidence to suggest that cognitive load is a key factor~\cite{lyell2017automation}. A recent survey on the use of AI by undergraduates suggests that 88\% of students have used AI tools for assessments~\cite{freeman2025student} and that around 10\% of ChatGPT users depend on it for a wide range of tasks~\cite{stojanov2024university}. At the same time many academic institutions and other organisations have policies that restrict the use of LLMs. There appears to be a similar tension in reporting on LLM use as was observed with smartphones, contrasting the risk of over-reliance with the limitless possibilities of a powerful new tool (and by extension the `out of touch' people who refuse to adopt the new technology)~\cite{10.1145/2470654.2466134}.

The potential for over-reliance is particularly concerning given the potential for CAIs to perpetuate existing bias and misinformation (informally `garbage in, garbage out'~\cite{mittelstadt2016ethics}), highlighting the importance of reliable information provenance~\cite{10.1145/3613905.3650862}). Even with neutral framing, LLM-powered search tools can significantly amplify users' pre-existing biases compared to traditional internet searches, potentially leading to increased polarization~\cite{10.1145/3613904.3642459}. Conversational search systems that reinforce existing biases exacerbate this problem, and even using LLM-powered search systems with opposing views has little effect on reducing this bias~\cite{10.1145/3613904.3642459}. Interestingly, however, it is also possible that AI-generated images could actually \textit{reduce} reporting bias by better reflecting content than images chosen by authors\cite{10.1145/3678884.3681907}. There is also the potential to generate new misinformation through hallucinations or malicious prompts~\cite{10.1145/3571730}. In response to this, researchers have begun to explore how CAIs themselves might be used to help detect misinformation for end users or human fact-checkers~\cite{10.1145/3686962, 10.1145/3613905.3648110, 10.1145/3626772.3657965}. This is particularly useful given that AI-generated misinformation tends to meet existing assessment guidelines designed to help people detect misinformation~\cite{10.1145/3544548.3581318} (e.g. by providing fictional citations, which can require more domain expertise to identify~\cite{10.1145/3726009, 10.1145/3184558.3188731}).

\section{Methods}
In order to answer the research questions above we developed a semi-structured interview protocol, with participants asked to complete two tasks using the copilot CAI integrated in the Edge web browser whilst verbalising their thoughts. We sought participants who were generally familiar with CAI and desktop web browsers but not the specific combination used for the study in order to reduce onboarding time and eliminate initial perceptions of CAI as a variable in the analysis. We therefore recruited students from the first author's university, who had experience with CAI but not Copilot. Participants were recruited using the first author's university's temporary work platform and received \pounds20 in compensation. Recruitment was done incrementally and we eventually settled on a participant pool of 20 as providing sufficient information power~\cite{malterud2016sample}. Our IRB approved all parts of the study.

We began by running three pilot interviews in order to judge the suitability of the tasks chosen, leading to adjustments to increase participant engagement with the topics of the research questions. Following this the first and second authors conducted the initial two interviews together to ensure familiarisation and alignment with the protocol, before conducting the remaining interviews individually. The stages of the interview are outlined below and the full protocol is included as supplemental material.

\subsection{Informed Consent}
Participants were sent an online consent form ahead of the study, and were given time to complete it before the interview if they had not already done so. They were invited to raise any questions with the interviewer before moving on.

\subsection{Familiarisation (approx. 10 minutes).} Each interview began with the interviewer introducing themselves and the study to the participant and discussing previous experiences they had had with CAI. Participants were then introduced to the current development release of Microsoft Edge. The browser has an integrated sidebar that allows easy use of the Copilot LLM. The chat-style interface makes it easy to get assistance from Copilot whilst browsing the web, including opening links suggested by the AI as well as using the AI to summarise or find information in the current tab. %(v118--v.0.2060.1) 
The interviewer opened Edge full screen on a laptop, showing the participant how to use the CAI and giving them a few minutes to familiarise themselves with the software and its functionality. To do so, the interviewer used the Copilot interface to get updates on the day's weather (``What's the weather in London today?'') and context dependant follow-up question on the UV index (``And what's the UV index [in London]?'') to demonstrate that the CAI 1) had access to real time information; and 2) could remember context from previous conversation turns. Participants were free to ask the researcher questions about how to use the software, but questions about how the CAI worked were not answered.

\subsection{Think-aloud Task Completion (approx. 30 minutes).} We then gave participants two tasks to complete using the copilot chat functionality. When designing the tasks we aimed for scenarios which would be encountered during typical usage but could be approached and answered in different ways. The first task was to find information on the health benefits and drawbacks of a cooking ingredient of their choice, and the second task was to plan a visit to a local city for the coming weekend, including transport, a trip to a local museum, and a nearby restaurant. In order to ensure that all participants used the CAI we asked them to start the task with it, before allowing them to browse onward from pages suggested by the CAI if they wished. Both tasks was designed to require information that participants were unlikely to know precisely beforehand but were likely to have a general sense of the bounds of a reasonable answer. Tasks were completed in the different browser windows to preserve the text of the chat, which was later saved and used to triangulate the analysis. Participants were asked to articulate their thoughts as they were completing the tasks, with the interviewer prompting them to continue articulating if they became silent. After completing each task we captured their beliefs and associated reasoning by asking ``why was this information in particular included?'', ``why were these linked websites chosen?'', and ``what other things might have been returned instead?'', all drawn from prior work on internet search~\cite{10.1145/3371390}.

\subsection{Follow-up (approx. 20 minutes).} Finally, we closed the interview by asking more general questions about participants' experience with the integrated CAI using questions from the literature on trust in technology~\cite{madsen2000measuring, cannizzaro2020trust} around the following constructs:
\begin{itemize}
    \item Reliability
    \item Understandability
    \item Technical Competence
    \item Faith in the answers given
    \item Privacy
\end{itemize}

\subsection{Analysis}
Interviews were recorded, transcribed, and cleaned by the researcher who gave the interview. Recordings were deleted after transcription. We performed thematic analysis following Braun and Clarke~\cite{braun2012thematic, braun2021thematic}, with the first and second authors developing and refining the initial codes together from a subset of the interviews. Coding was then done iteratively with frequent meetings to discuss the data and edge cases, refining the codes as necessary. The authors then collaboratively developed the themes reported in the next section. In line with established norms for thematic analysis, we do not report inter-rater reliability scores that suggest a single truth exists in the data~\cite{braun2021thematic}. We also echo the push to move away from applying quantitive standards of generalisability to qualitative research, instead we focus on transferability by including relevant details about the context, participants, and circumstances of the study throughout the paper to allow the reader to evaluate the potential for applying the results to other settings~\cite{braun2021thematic}.

\section{Results}
Participants were aged between 19 and 51, with an average age of 25.3 ($\sigma = 8.4$), with 17 identifing as women and three as men. Participants had experience with a range of non-copilot CAIs, listed in Table~\ref{tab:participants-CAI}. Below we report the four themes generated during the analysis around participants' beliefs when using Copilot.

\begin{table}
    \centering
    \begin{tabular}{|r|l|}
    \toprule
    Participant & CAI used \\
    \hline
    P1 & ChatGPT, Google Assistant \\
    P2 & Siri \\
    P3 & ChatGPT, Notion (integrated GPT-4) \\
    P4 & ChatGPT, Bard \\
    P5 & ChatGPT, Bard, Bing AI \\
    P6 & ChatGPT, Siri \\
    P7 & ChatGPT, Siri, Alexa \\
    P8 & Siri, unspecified LLM \\
    P9 & ChatGPT, Siri \\
    P10 & Siri \\
    P11 & ChatGPT \\
    P12 & ChatGPT, Alexa, Google Assistant, Bard \\
    P13 & ChatGPT, Siri \\
    P14 & ChatGPT, unspecified LLM \\
    P15 & ChatGPT, Jasper \\
    P16 & ChatGPT \\
    P17 & ChatGPT, Snapchat (integrated GPT model) \\
    P18 & Gemini, Google Assistant \\
    P19 & ChatGPT, Gemini \\
    P20 & ChatGPT, Bard \\
    \bottomrule
    \end{tabular}
    \caption{Participant experience with CAIs.}
    \label{tab:participants-CAI}
\end{table}

\subsection{Belief 1: I Know How to Use LLMs (RQ1)}
While almost every participant had individually developed a strategy for writing prompts, it was striking that there was no dominant approach across the interviews and that strategies were often conflicting. When asked why/how they had adopted their strategy of choice, we found that these were tied to underlying intuitions about how Copilot generated answers internally (Table~\ref{tab:prompting}). This fragmentation of prompt strategies appeared to be driven by the informal and ad-hoc way that participants discovered different ways of using CAI. The main source of these insights was trial and error through personal experience---using one or more CAI and adapting when they were unable to get the desired results---but also included friends and social media. No participants mentioned using material provided by the CAI vendor to more effectively construct prompts.

\begin{table}
    \centering
    \begin{tabular}{|l|l|}
        \toprule
        Strategy & Belief \\
        \hline
        Being as descriptive as possible & A more detailed prompt generates a more detailed answer \\
        \hline
        Start broad and refine & \multirow{3}{*}{Details are missed with longer prompts} \\
        Divide the question/task into several short prompts & \\
        Being direct and concise & \\
        \hline
        Ask it like asking a person & The system responds like a person \\
        \bottomrule
    \end{tabular}
    \caption{Prompt strategies and corresponding beliefs about Copilot.}
    \label{tab:prompting}
\end{table}

\begin{quote}
\textbf{P07}: \textit{I see them online and all over social media, it's like, hey, if you want this, give ChatGPT this prompt. So instead of just saying, hey, read my paper, you got to tell them, assume you're an expert on XYZ and then read it. So it's just a better prompt.}

\textbf{Interviewer}: Right, so have you ever searched for a prompt then to use?

\textbf{P07}: Oh yeah, multiple times
\end{quote}

Perhaps because of this, participants frequently blamed themselves during the tasks when Copilot didn't return the information they were looking for; if only they had written a prompt that was more specific, more defensive, or otherwise `better', then Copilot would have given them what they wanted. The uncertainty around how best to phrase questions resulted from a feeling that phrasing was important, and that asking the same question in a different way would generate a different answer. In contrast to this, there was a common belief that asking the same question at a future point in time would yield the same response, so long as the sources that copilot used to gather data on the question remained the same (i.e. the data it had access to on that particular topic).

\begin{quote}
\textbf{P08}: \textit{I think it would depend it would depend a lot on my questions, how I formulate them maybe.} 

\textbf{Interviewer}: Oh, so asking the same question but in a different way would come back with a different answer? 

\textbf{P08}: \textit{Yeah, maybe. I would think it depends depends a lot on the way I asked the question. Maybe if I asked the questions exactly the same way it would come up with the same answers, or maybe if I asked them a different way it would come up with different answers} 
\end{quote}

In more extreme cases, the varying quality of responses meant that they were seen as something of a lottery: ``\textit{That's a random... oh my God, that's a random coffee shop in the middle of Belgium!}'' [P03], ``\textit{Yeah, I mean sometimes I put stuff in and it's just completely having a bit of a moment where I don't really know what it's talking about, but that's quite rare}'' [P15].

\subsection{Belief 2: I Know I Need to Fact Check (RQ1, RQ2)}
The most common conceptualisation of the LLM was as a search engine, something that would take a prompt, run a web search on that prompt, and then summarise the results: ``\textit{I think what it's done is it's made its own search and then it's given me a concise list of this search}'' [P05]. This is surprisingly close to how the Copilot works behind the scenes, utilising retrieval-augmented generation techniques to ground the LLM's responses in the results of a Bing search. It also echoed the pre-responses shown by Copilot after users entered a prompt but before the full response was shown (e.g. ``searching for vegetarian restaurants near Soho''). We return to this point in the discussion.

When asked what metrics Copilot used to decide what information to returned, there was widespread consensus that results were primarily chosen based on popularity. However, when pressed further there were a range of different understandings about what that might mean in practice:
\begin{itemize}
    \item The top results as ranked by a search engine (``\textit{I mean this is like the top 5 searches in Google, right?}'' [P17])
    \item The web pages with the most traffic (``\textit{I'm assuming that they have the data from the amount of clicks [..] as the first website, that would have the most amount of clicks}'' [P03])
    \item Information that occurred most frequently across all of its sources (``\textit{I'm sure that [in] everything that is written about honey, this is a common point}'' [P20])
    \item Previous searches or Copilot interactions (``\textit{Other people have asked the similar question related to my question and they gave response to the AI like this is a good answer or this is a bad answer and it can self-teach itself to tell the customer this is the popular answer or this is the answer that most people don't like}'' [P14])
    \item Proximity, for places and events (``\textit{I just think it just went through locations which are together and checked if they are in that walking distance and suggested me one of them}'' [P18])
\end{itemize}

%\subsection{Truth, Trust, and Relevance (RQ2)}\label{sec:results-truth-trust-relevanvce}
There was less agreement, however, around the how to evaluate Copilot's answers. Beyond the potential for hallucinations, which was well known, participants also talked about the ability of Copilot to discern the authority or veracity of sources. It was often felt that the results given by Copilot \textit{should} be accurate and from authoritative sources, but on reflection there was a lack of certainty about whether this was even possible given the limits of the technology: ``\textit{I'm not really sure how the actual AI would know if a link is like reputable or not. Or like, if the opinion on that site is trusted.}'' [P13]. This reinforced a common sentiment amongst participants that results that had real world consequences would need to be fact-checked before being acted upon.

Interestingly, in many cases the go-to sources of information to verify the output of the LLM were the same search engines that participants believed that Copilot was using to generate its results: ``\textit{If it looks like interesting, I'll go and look deeper in the traditional way, Google [..] Not just, oh, okay, ChatGPT told me this, so I'm going to do it.}'' [P19]. This raises interesting questions around the efficacy of checking claims using the same sources as the LLMs themselves that we explore more in the Discussion. On the other hand, some participants avoided this problem by using LLMs to satisfy curiosity or generate ideas in a way that didn't require full confidence in the results: ``\textit{Basically, I use it just to get ideas, mainly for [university], I'm studying my masters here. So I don't use it as my principal source of information. I like to search for ideas, and then after getting those ideas, I go deeper into different other more reliable sources}'' [P19].

\subsection{Belief 3: I Know What to Be Cautious Of (RQ2)}
The common conceptualisation described above of the LLM as a sophisticated search engine meant that participants imported their existing heuristics for those products. Participants were often cautious about subtle or undisclosed prioritisation of particular URL citations by advertising partners, similar to the in-feed or pre-results ads that they had experienced on other platforms: ``\textit{I know you can pay to sponsor your website online. So when someone Googles example holidays in Europe, yours would come up. I don't know if these websites are either paying just to somehow have their answers come up on [Copilot]. Yeah, I don't know if that's even a thing to advertise on to this, but I wouldn't be surprised}'' [P07]. This line of reasoning was often triggered by results that the participants felt was incongruent with how they expected the CAI to respond and had difficulty explaining, and seemed to utilise the same skills that they had developed to navigate other online resources. This defensive approach to advertising is likely based on poor disclosure practices observed on social media platforms~\cite{asa-influencers-2024}.

\begin{quote}
\textbf{P15}: \textit{I kind of find because with Google sometimes they have sponsored posts come up first and also I think it's just faster using Tik Tok as opposed to Google, so videos are shorter.}

\textbf{Interviewer}: Do TikTok or Instagram not have sponsored posts?

\textbf{P15}: \textit{Yeah they also do, you have to check like a lot of these people who are endorsing these restaurants and stuff also got a free meal out of it so the same applies.}
\end{quote}

Similarly, when asked why Microsoft-affiliated links had been returned as citations many participants suggested that Copilot would be programmed to favour in-house sources. Unlike paid advertising, which often carried with it negative connotations about the honesty or truthfulness of results, prioritising Microsoft links was treated much more neutrally: ``\textit{This is the Microsoft AI, right? So they included their own source because in the end, as a business, your main motive is to make sure you're seen everywhere [..] that's just basic business}'' [P17].

For matters of fact, URL citations were implicitly trusted by participants regardless of the website they pointed to. Their inclusion in results allayed general concerns about the accuracy of the results returned by copilot, reduced worries around hallucinations, and was compared favourably to other CAI that the participants had used which had not included them. The academic environment from which our participants were recruited places an implicit trust that citations in research papers are accurate and support the claims they appear by in-text.

``\textit{I like the idea because I want to see if the information the AI provided has actual words and actual data I can trace and give me something like, you're not making things up for me to fool me}'' [P14].

In most cases the presence of the citations alone was enough to establish the credibility of the information to the point where the participant felt they didn't need to investigate further, especially when things seemed plausible: ``\textit{I think for me I would not go the extra step and look at the [citation] website because if it's giving me the information that I need, I don't want to spend more time on it}'' [P16]. Citations were often thought to be the most relevant source drawn on by co-pilot, despite the fact when participants did investigate further they sometimes found the links to be unhelpful, irrelevant, or misleading. As we discuss in Section~\ref{sec:discussion}, citations have academic connotations and it is likely that every other time they have been experienced by participants they have been accurate and made with good intentions (unlike, for example, with advertising disclosures).

\begin{quote}
[P04 asks for drawbacks of cooking with thyme]

\textbf{P04}: Yeah, usually links with, like, .org, org.uk, they are usually more legitimate.

\textbf{Interviewer}: So it's going to be more reliable than Wikipedia?

\textbf{P04}: Yeah, it's quite... quite straightforward. Let's try this. Yeah, it's from NHS Foundation Trust, so I would assume it's...

\textbf{Interviewer}: What is this website actually about, though?

\textbf{P04}: Oh, this is... wait, this is... I think they're talking about acronyms and not the.

\textbf{Interviewer}: ``Structure conversations with a person in distress''

\textbf{P04}: This is something different, I think. Yeah, so it's not... this is definitely correct [points to another citation], but this one is off-topic [\url{www.sageandthymetraining.org.uk}]. Yeah, it's... mnemonic and not the actual product, yeah.

\textbf{Interviewer}: So I guess how does that change how you think about the information that's come back?

\textbf{P04}: Well, I'm actually confused because this is correct, but they linked it to... Oh, sorry, this... These are correct, but they link it to this page, which is nothing to do with the benefits or the drawbacks of thyme.

\textbf{Interviewer}: Right. And when you say correct, what are you basing that on just to help me figure out what you mean?

\textbf{P04}: I guess on how logical it is, like they're not writing something completely off-topic. I could imagine that excessive bleeding or bruising would be something possible as a drawback. Logically, okay. Yeah, but I guess the fact that it's not... the website is not related might make me think, like, oh, is this true or not?
\end{quote}

When asked whether they trusted Copilot to use their data appropriately, participants were largely unconcerned regardless of whether they had previously considered this or whether they believed that the data collected would stay within Microsoft as a company. The only scenario with significant negative sentiment was the use of collected data for advertising. There were no special risks highlighted beyond the background risk of a data breach that was seen to come with using any online service or connected device. Occasionally participants described this as a trade-off between privacy and convenience that came down heavily on the side of convenience.

\subsection{Belief 4: I Know What AI Looks Like (RQ1)}
All participants considered Copilot to be a clear example of AI, but were often puzzled when asked how AI might be used on other platforms they were familiar with to do similar tasks (e.g. ``how might AI be used when you're looking for cafes on Instagram?''). This line of inquiry revealed several different conceptualisations of how the concept of AI related to LLMs. For P19, AI and LLMs were the same thing, while others saw it as a wider phenomenon but still one that was fundamentally \textit{new}. Except for the few who had prior knowledge of machine learning, participants were vague about potential other uses of AI. A third conceptualisation that emerged was that AI was related to certain devices and use-cases: ``\textit{the AI right now, it's, like, on my phone, not on my laptop}'' [P17]. In general, AI was considered a different thing than the algorithms that participants described operating elsewhere, such as on social media platforms.

Learning was an important component of this, with a common complaint about Copilot and its recommendations (and to a lesser extent other CAI that participants had used) being that it didn't `know them' well enough to provide good recommendations compared to what participants were used to on other platforms. If they wanted to find something, then they needed to provide all of the details via their prompt. In comparison, social media platforms managed to achieve similar results without any conscious effort: \textit{``I think with this, what's it called? Co-pilot? You can make it personalised, but it depends on what you plug into the chat box [..] whereas with TikTok obviously it's got an algorithm, so it knows what videos I interact with it knows what places I follow [..] so I think with this you have to plug in the information for it to be personalised to you whereas Tik Tok it's more of an algorithm, it kind of does it for you''} [P15]. Combined with the above conceptualisations of AI, this shows how design and interaction modality have a significant influence on the `visibility' of machine learning features.

Another dimension to this was the `coolness' of the answers given to them by Copilot. Several participants felt that while Copilot was able to find more established attractions and events, it paled in comparison to other platforms such as TikTok when it came to `trending' things to do. ``\textit{P13: I guess so. I think especially TikTok, it's very like, sort of like what's trending based. So, it will always be things that like photograph good or like video well to be able to make like engaging content. Whereas if you ask the like this chat thing for things to do this weekend, it will probably [..] give you things where like there's loads of things written about [them], like the British Museum}'' [P13]. Copilot wasn't believed to be drawing on these social networks as sources, and as a result was limited to slower and less trendy (but still real-time) written content on the web. Some participants directly associated this with the different algorithms used to surface content on social media and web search, which we discuss more in the next section.

%It is worth noting that this accurately reflects the current operation of Copilot in Edge, which uses a combination of URL, page title, user's prompt, current conversation history, and browsing context of the current session to generate responses.\footnote{\url{https://learn.microsoft.com/en-us/copilot/edge}} Previous conversations are not utilised. Another relevant factor was the human-ness of the Copilot interface, as evidenced by the ``talk to it like a person'' prompt strategy. Prior work has shown how human-likeness in a system's interface leads to increased performance expectations; if the system appears more human then people expect it to perform to the level that a human would.

\section{Discussion}\label{sec:discussion}
\subsection{How Do We Design For \textit{Untrustworthy} AI?}
LLMs are often marketed as tools designed to reduce the time and effort required to complete tasks, with the caveat that their output should be verified before relying on it. But how to do that is an open question. Applying dual process theory to Beliefs 2 and 3, we can see how the citations functioned as low-effort (system 1) markers of veracity, which is likely why they were not investigated further. On the other hand, when prompted about trustworthiness or when describing high-stakes tasks, participants described higher effort (system 2) fact-checking methods that they already used (mostly web search).

This challenges the idea that citations can provide a reliable transparency mechanism that mitigates concerns around hallucinated or incorrect results. Reviewing the Copilot inputs and outputs we found that spurious footnotes were just as likely to appear plausible as errors in the main response text. This suggests that LLM-generated citations may never be able to mitigate problems about the accuracy of LLM outputs, as if the LLM could make a mistake in the main response then it could make the same kind of mistake in the accompanying citation~\cite{10.1145/3708359.3712160}. If users take the presence of the citation as evidence of a correct answer then they do not even offer another opportunity to spot discrepancies in a response that might need further verification. 

The probabilistic nature of LLMs is not a temporary limitation to be engineered away, but a fundamental characteristic that demands a shift in how we design conversational AI. Current approaches such as retrieval-augmented generation, citations, or small-print disclaimers urging users to verify outputs treat fallibility as a surface problem to be papered over rather than a core constraint to design around. As our findings demonstrate, these mechanisms can backfire: citations become system 1 markers of veracity rather than invitations to scrutinise, and RAG systems can hallucinate sources as readily as facts. If we accept that errors are inevitable rather than exceptional, the design question shifts from \textit{``how do we make users trust this system?''} to \textit{``how do we help users recognise when this system is wrong?''} This reframing moves us from designing for \textit{trust despite fallibility} toward designing for \textit{appropriate skepticism because of fallibility}.

What might this look like in practice? We saw participants reach for the same tools to verify answers that LLMs use to generate them: search engines. This strategy probably came to mind as the process that participants used to find information online outside of LLMs, so why not include a button that would open the current prompt in a search engine or display search results alongside LLM output? This is a small enough action that it could be invoked via system 1, vastly increasingly the chance of it being used. Of course, this approach is only as good as a user's ability to identify sources that are accurate and appropriate. An alternative would be to include confidence markers in the LLM's output that could indicate via system 1 that the result might not be trustworthy---engineering for productive doubt rather than scaffolding confidence. 

Finally, with evidence suggesting that LLM search tools can amplify biases compared to traditional web searches~\cite{10.1145/3613904.3642459}, it was interesting to see one participant comment on the way that Copilot appeared to take a side when reporting current events, which was exacerbated in one case by the CAI refusing to `admit' taking a stance. \textit{I think it wants to tell us that he's not doing well, but then he..  the AI is not allowed to give his own opinion, so he just sneaks in an article that might reflect it’s opinion''} [P07]. This shows how being \textit{perceived} as neutral can be an impossible task, particularly when relying on existing articles that will have a political leaning. At the very least, CAIs should provide citations for news articles so that users can consider this context when interpreting responses.

\subsection{How Design Shaped Perceptions}\label{sec:design-mental-models}
There was an interesting tension where participants compared Copilot with internet search engines as well as social media, with the CAI inhabiting a space between the two. Alongside prompt strategies from other LLMs, participants used web search analogies to explain where Copilot was sourcing and prioritising information. In the tasks, we saw many concepts frequently present in mental models of web search~\cite{10.1145/3371390}, including popularity metrics (e.g. most clicks, most traffic), commercial interests (e.g. advertising, payment for prominence), location, and authority. This makes sense given the similarity between contexts, as well as the search related cues given to users by Copilot---prompts were followed by text bubbles saying e.g. ``searching for vegetarian restaurants near Soho''. In terms of social media, in comparisons with platforms like TikTok and Instagram it was suggested that these would have provided suggestions for places, experiences, and events that would have been more `real' or `trending' (and thus preferred). This was paired with a characterisation of the web (and therefore search engines) as being too slow, reinforced when Copilot struggled to offer recommendations for specific time periods like events happening this weekend. Within their social media comparisons participants made a distinction between algorithms (used by participants to mean what is used to select content on social networks) and AI (taken to mean CAI). The relationship between LLM CAI design cues and subsequent perceptions is a fascinating one that is worth unpacking further in future work, as it may also influence the repurposing of knowledge and strategies that people have developed for existing platforms and services; people tend to have different conceptualisations of risk for search engines and social media and regulate their usage of them as a result.

%The importance of query wording was also identified in the interview themes but focused on the interpretation of the words by Copilot rather than how they were used to select retrieved content. Participants often described the range of apps and services they would use to complete different tasks (e.g. Google Maps for navigation and Instagram for finding places to eat), mirroring prior work showing how trust levels vary based on what content is being sought~\cite{10.1145/3613905.3650862}. This suggests that the most useful and most easily understood architecture of an LLM might be one where the LLM composes outputs from existing tools (e.g. web search, mapping directions) rather than producing output content itself.

%Several participants acknowledged the skills they needed to develop in order to assess the authenticity of the content they provided. Future CAIs may incorporate increasing quantities of conversation history and other usage data to provide more personalised results. However, at present, this can be limited by the number of tokens that the foundation model will process.

%The readiness with which participants associated the Copilot with the concept of `AI' and the difficulty that many had talking about AI outside of CAIs highlights the nebulous use of AI as a cultural, marketing, and ideological term. 

\subsection{Data and Sharing}
Observing across the interviews, we saw that participants provided a large amount of information to Copilot very quickly (either directly or in a way that could easily be inferred). Given that concerns about how this data would be handled were like those expressed for adjacent technologies, it is likely that these disclosures were not entirely conscious choices. It is not clear the extent to which this happened because LLMs as a class of tool are new and participants had not learned how to use them according to their privacy preferences, or because conversational interfaces inherently encourage disclosure (and can be adjusted to increase the rate of information extraction~\cite{zhan2025malicious}).

We did see evidence that participants tailored their approach to using Copilot as if it were ``\textit{a conversation, like if there was a person in front of me}'' [P06]. Indeed, unlike for web search, many felt that providing as much information as possible to Copilot was the best strategy for finding information, without mentioning associated privacy considerations. This sometimes prompted reflection when participants were specifically asked about privacy issues, such as when P15 realised that asking for halal restaurants would allow Copilot to infer information about their religion. This is particularly relevant given that LLMs potentially have the cultural context required to be able to make these inferences without additional training. Making shared and inferred information visible through mechanisms similar to those used for targeted advertising would serve the dual purpose of highlighting to users what the system knows about them as well as providing an opportunity for deletion and correction (two key data rights).

Unlike for adjacent technologies such as voice assistants, it does not appear that participants' trust in the software they were using was anchored in their previous relationship with the company that developed it~\cite{10.1145/3479515}. Many participants described having prior experience with other LLMs, particularly ChatGPT, but none said anything that would suggest a different relationship with OpenAI---a company with which they would have had no past interactions---than with Microsoft, who provide much of the software used at the university. What was also striking was that during the study none of the participants ever asked Copilot how it generated its responses nor what would happen to the data that was given to it. This suggests that transparency information about the data a system collects might need to be signposted in other ways.

\section{Limitations}
As mentioned above, we had to balance participant's previous experiences with LLMs having too little experience with LLMs in general would have dramatically increased familiarisation time and negated our ability to conclude the LLM's design, as we would have been observing first reactions to LLMs more broadly. To preserve the ecological validity of the setup used by participants, we were also constrained to commercially available software with an integrated LLM, of which there was a limited selection at the time of the study.

This led us to the decision to recruit from our university and to use Copilot, as students have generally had experience with LLMs but not with Copilot and by using OpenAI's GPT4 it was representative of the wider market for CAIs/LLMs. At the same time this meant that the participant pool was skewed in terms of age and gender identity. Finally, as the study was not focused on any specific aspects of Copilot, we did not control the specific (sub)version of Copilot that was used during the interviews. This may have meant that the experience for some participants was slightly different than for others, but this is unlikely to have affected the overall results of the study.

\section{Conclusion}
In conclusion, this study explored how users perceive and conceptualise CAI integrated into existing software products. Through the interviews where participants interacted with a CAI built into a web browser we explored their mental models of the CAI, often considered akin to an advanced search engine. Participants also used a diverse array of prompt strategies that were largely discovered ad-hoc, each of which was tied to a particular belief about how the CAI functioned. Turning to the perceived strengths, weaknesses, and risks of this technology, we saw how citations included in copilot responses were seen as a clear trust signal, although this continued even when citations were shown to be spurious. Participants were wary of the inclusion of advertising within Copilot, but more tolerant of favoured first party links. In general, the CAI was considered to very visibly \textit{be} AI, highlighting a shift towards LLM-based technologies being the public face of what is in reality a vast field.

Discussing these findings in the context of prior work in the area, we consider the implications of participants disclosing large amounts of information to Copilot without realising, as well as the cultural positioning of Copilot in relation to other popular AI-driven platforms that our participants had encountered. Looking forward, we use these insights to suggest a pivot in the way that explanations are used in CAIs as indicators of system errors, and a framing of these technologies as `untrustworthy AI' in contrast to existing initiatives.

\begin{acks}
This work was supported by the Engineering and Physical Sciences Research Council through grant EP/T026723/1 and by a PROMETEO 2024 (CIPROM/2023/23) grant of Conselleria d’Educació, Universitats i Ocupació (Generalitat Valenciana). 
\end{acks}

\bibliographystyle{ACM-Reference-Format}
\bibliography{main}

\appendix
\end{document}